\documentclass[fleqn,usenatbib]{mnras}
\usepackage[british]{babel}

\usepackage[T1]{fontenc}
\usepackage{ae,aecompl}

\usepackage{graphicx}	% Including figure files
\usepackage{amsmath}	% Advanced maths commands
\usepackage{amssymb}	% Extra maths symbols
\usepackage{mathptmx}
\usepackage{txfonts}
\usepackage{subcaption}
\volume{{\rm in press}}

\defcitealias{2009AJ....138..402S}{Paper I}
\defcitealias{2012AJ....143..149S}{Paper II}
\defcitealias{2015MNRAS.452.2858K}{Paper III}

\title[Survey for CSPNe I. Methods and First Results]{A Survey for hot Central Stars of Planetary Nebulae I. Methods and First Results}

\author[G. Kanarek et al.]{Graham C. Kanarek,$^{1,2}$\thanks{E-mail: gray@astro.columbia.edu (GK)}
Michael M. Shara,$^{2}$
Jacqueline K. Faherty,$^{2,3,4}$
\newauthor
David Zurek $^{2}$
and Anthony F.J. Moffat$^{5})$
\\
$^{1}$Columbia University, 116th St \& Broadway, New York, NY 10027, USA\\
$^{2}$American Museum of Natural History, 79th Street and Central Park West, NewYork, NY, 10024-5192, USA\\
$^{3}$Hubble Fellow\\
$^{4}$Department of Terrestrial Magnetism, Carnegie Institution of Washington, \\
5241 Broad Branch Road NW, Washington, DC 20015, USA\\
$^{5}$D\'{e}partment de Physique, Universit\'{e} de Montr\'{e}al, CP 6128 Succ. C-V, Montr\'{e}al, QC, H3C 3J7, Canada
}

\date{Accepted XXX. Received YYY; in original form ZZZ}

\pubyear{2015}

\begin{document}
\label{firstpage}
\pagerange{\pageref{firstpage}--\pageref{lastpage}}
\maketitle

% Abstract of the paper
\begin{abstract}
We present the results of initial spectrographic followup with the Very Large Telescope (UT3, Melipal) for $K_s\ge14$ Galactic plane \ion{C}{iv} emission-line candidates in the near-infrared (NIR). These 7 faint stars all display prominent \ion{He}{i} and \ion{C}{iv} emission lines characteristic of a carbon-rich Wolf--Rayet star. They have NIR colours which are much too blue to be those of distant, classical WR stars. The magnitudes and colours are compatible with those expected for central stars of planetary nebulae, and are likely to come from massive progenitor populations, and themselves be more massive than any sample of planetary nebulae known. Our survey has identified thousands of such candidates.
\end{abstract}

% Select between one and six entries from the list of approved keywords.
% Don't make up new ones.
\begin{keywords}
planetary nebulae: general -- infrared: stars -- stars: emission line, Be -- Galaxy: disc -- techniques: spectroscopic
\end{keywords}

%%%%%%%%%%%%%%%%%%%%%%%%%%%%%%%%%%%%%%%%%%%%%%%%%%

%%%	%%%%%%%%%%%%%% BODY OF PAPER %%%%%%%%%%%%%%%%%%

\section{Introduction}
The Galactic plane, shrouded in dust, remains a difficult region of the sky to probe. While optical surveys are limited to distances of a few kpc, exploration of the Milky Way in the near-infrared (NIR) range suffers much less from dust extinction. As the number of high-quality, ground-based NIR instruments grows, so too does our ability to probe the most distant reaches of the Galaxy. Planetary nebulae (PNe) have strong emission lines in the $K_s$-band NIR spectrum, most notably \ion{He}{i} (2.058 \micron) and the \ion{H}{i}--Br-$\gamma$ doublet (2.166 \micron) \citep{2006AJ....131.1515L}, and are often identified during Wolf--Rayet (WR) surveys \citealp[cf.][]{2012AJ....143..149S,2015MNRAS.452.2858K}. The central stars of planetary nebulae (CSPNe) are far more rarely identified in continuum surveys due to their intrinsic faintness. Galactic plane CSPNe are heavily reddened, so narrowband NIR surveys are essential to find large samples.

Of the currently $\sim500$ Galactic CSPNe with spectroscopic classification \citep{2011A&A...526A...6W}, approximately 30 per cent are H-poor, with $\sim22$ per cent displaying WR-like emission lines. These have been named [WR]s, the notation first introduced in \citet{1981SSRv...28..227V}. The great majority are carbon-dominated [WC]s showing strong carbon emission, with the first [WN] recently confirmed in \citet{2012MNRAS.423..934M}. The hydrogen deficiency and carbon-richness of these [WR] CSPN spectra is likely due to dredge-up from a late-stage thermal pulse \citep{2006PASP..118..183W}. Nebulae associated with early-type [WC] stars are typically more evolved than those of late-type [WC]s, implying an evolutionary sequence from late to early type \citep{2001A&A...367..983P}; this is in contrast to classical WR stars, which show no obvious intra-type evolution \citep{2007ARA&A..45..177C}.

In \citet[Paper I]{2009AJ....138..402S}, \citet[Paper II]{2012AJ....143..149S}, and \citet[Paper III]{2015MNRAS.452.2858K}, we presented results from a large-area Galactic plane survey in the $K_s$-band, searching for emission-line objects in crowded fields. While our prime targets are new and distant classical WR stars, we also find planetary nebulae and other hot stars. The majority of candidates from this survey are at $14 \le K_s \le15$, and have not yet undergone spectroscopy, as they require significant time on large telescopes equipped with sensitive NIR spectrographs.

In this paper we present, as a proof of concept, our first results from followup spectroscopy of very faint, $K_s$-band \ion{C}{iv} emission-line targets, using the Melipal (UT3) telescope, part of the Very Large Telescope (VLT) at Paranal. This proof of concept survey has led to the identification of 7 likely [WC]s. In sections~\ref{sec:cand}, \ref{sec:obs}, and~\ref{sec:redu}, we discuss the candidate selection, observations, and data reduction, with the results presented in section~\ref{sec:res}, and a discussion of the new [WC]s in context in section~\ref{sec:mass}. We summarize our conclusions in section~\ref{sec:awk}.

\section{Data Collection}\label{sec:data}
The survey first described in \citetalias{2009AJ....138..402S} consists of infrared imaging observations carried out on the SMARTS 1.5m telescope at the Cerro Tololo Inter-American Observatory (CTIO) over 200 nights in 2005-2006. The survey involved images taken through 4 narrowband (NB) filters, centered on four of the most prominent emission lines in the $K$-band WR spectrum \citep{1997ApJ...486..420F}, as well as two continuum filters in the same band, with the purpose of identifying WR candidates for spectrographic confirmation. The survey covered 300~deg$^2$ of the Galactic plane, from $-90^{\circ} \le l \le +60^{\circ}$ and $b \pm 1^{\circ}$. Using the images from this survey, in \citetalias{2009AJ....138..402S,2012AJ....143..149S,2015MNRAS.452.2858K}, as well as \citet{2014AJ....147..115F}, we have discovered $\sim25\%$ of the currently-known Galactic WR population.

\subsection{Candidate Selection}\label{sec:cand}
\begin{figure}
\includegraphics[width=\columnwidth]{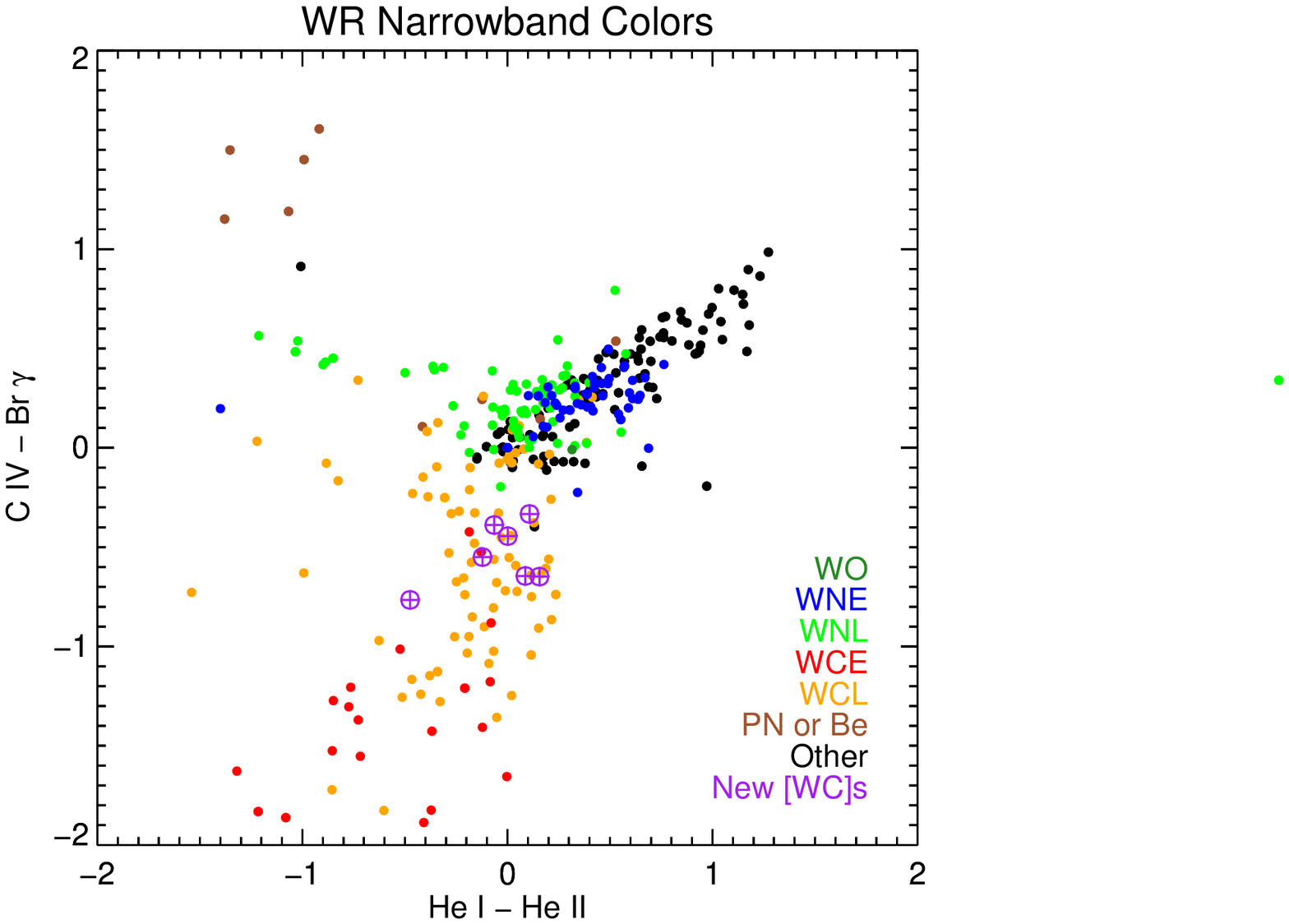}
\caption{Narrowband colour-colour plot, using emission-line filters in the NIR. WCs separate out clearly, and the 7 new objects presented in this paper (coloured purple in the figure) are consistent with objects which have \ion{C}{iv} emission lines. "Other" objects (coloured black) are a mix of red supergiants and young stellar objects with weak Br-$\gamma$ emission, and some non-emitting red giants.}
\label{fig:nbcolors}
\end{figure}

When reducing the images for this survey, we calculated magnitudes for every object imaged in each NB filter using \textsc{idl-daophot}'s \textsc{aper} aperture photometry routines, and then scaled the magnitudes from each NB image to match the \textit{2MASS} $K_s$ filter \citep{2006AJ....131.1163S}, for those objects which had analogues in the \textit{2MASS} point-source catalogue. Figure~\ref{fig:nbcolors} shows a NB colour-colour plot for spectrographically-characterized emission-line objects of various types (primarily WR stars and PNe).

\begin{figure}
\includegraphics[width=\columnwidth]{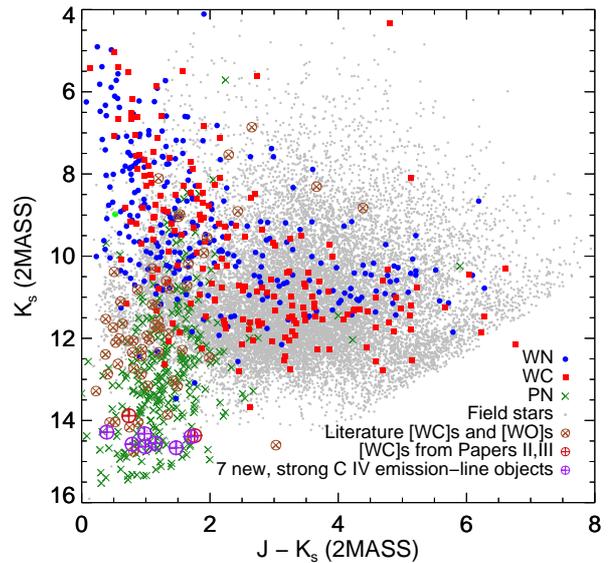}
\caption{\textit{2MASS} $J - K_s$ vs $K_s$ for field stars, WNs and WCs (in the literature), PNe, and [WC]s from the literature, Papers II \& III (see section~\ref{sec:oth}), and the candidates presented in this paper. The new objects' positions in the infrared color-magnitude diagram strongly suggests that they are hot central stars of planetary nebulae. They are too blue, given their faint $K_s$ magnitudes, to be classical WR stars.}
\label{fig:kjk}
\end{figure}

Since our goal is to isolate [WC] stars, we chose candidates that \textit{simultaneously} display CIV emission, \textit{and} are too faint and blue to be classical WR stars in a $K_s$ versus $J-K_s$ colour-magnitude diagram (see Figure~\ref{fig:kjk}). The \ion{C}{iv} emission-line condition is fulfilled by imposing the constraints \ion{C}{iv}$-$Br-$\gamma\le-0.3$ and \ion{He}{i}$-$\ion{He}{ii}$\le -0.3$ (see Figure~\ref{fig:nbcolors}). The color-magnitude constraints are $14\le K_s \le 15$ and $0 \le J-K_s \le 2.0$. (see Figure~\ref{fig:kjk}). The latter constraint excludes classical WR stars, and places our candidates solidly in the regime of known CSPNe and literature [WC] and [WO] stars. The seven objects presented in this paper are $\sim2$ magnitudes fainter than the classical WR stars of similar colour, indicating much less extinction, and therefore much lower luminosities, than one would observe in classical WR stars at the same magnitude.

\subsection{Observations}\label{sec:obs}
Spectrographic confirmation of our faint emission candidates requires an 8-meter-class telescope equipped with an IR spectrograph. In the summer of 2013, we observed 7 strong emission-line candidates, selected to be likely [WC] stars, with the Infrared Spectrometer And Array Camera \citep[ISAAC][]{1998Msngr..94....7M} on the VLT. We used the instrument in SWS1-MR mode, with a 0.6 arcsec slit, for a resolution of $R=4400$ over the range 2.049-2.175 $\mu$m. Each target was observed for 240 seconds per dither, with at least 6 dithers for each star. Finder charts for these candidates are shown in the appendix, Figure~\ref{fig:fc}.

\subsection{Reduction}\label{sec:redu}
The raw data were reduced via a \textsc{python} script, making use of the \textsc{python-cpl}\footnote{https://pypi.python.org/pypi/python-cpl/0.6} package to interface with the ESO Common Pipeline Library (CPL). Each object was processed with the \textsc{isaac\_spc\_flat}, \textsc{isaac\_spc\_arc}, \textsc{isaac\_spc\_startrace}, and \textsc{isaac\_spc\_jitter} recipes from the CPL. Then, a smoothing algorithm, which increases the smoothing length in areas of high noise (which has the effect of preferentially smoothing noise more than signal), was applied to improve the spectrum quality.

\section{Results}\label{sec:res}
\begin{figure}
\includegraphics[width=\linewidth]{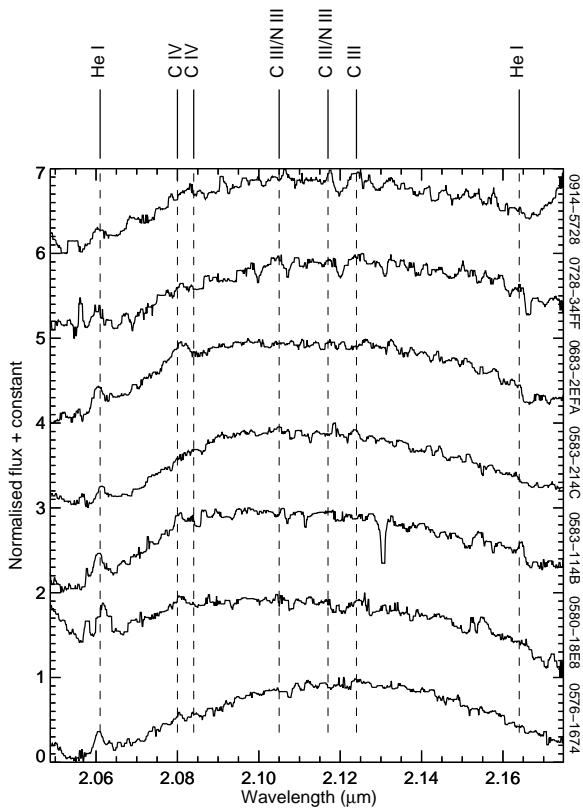}
\caption{All new [WC] stars identified in this paper. While the S/N of each spectrum is modest, it is sufficient to see the strong emission features at 2.059 $\mu$m (\ion{He}{i}) and 2.08 $\mu$m (\ion{C}{iv}). The ``absorption'' at 2.124~$\mu$m in spectrum 0583-114B is an instrumental artifact.}
\label{fig:specs}
\end{figure}

\begin{figure}
\includegraphics[width=\columnwidth]{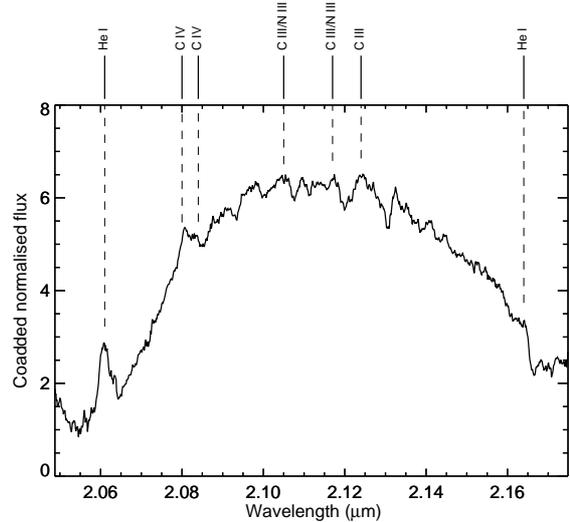}
\caption{All 7 spectra from Figure~\ref{fig:specs}, normalised and coadded. The 2.07 $\mu$m \ion{He}{i} and 2.08 $\mu$m \ion{C}{iv} lines are now obvious, as is the \ion{C}{iii} line at 2.122 $\mu$m.}
\label{fig:coadd}
\end{figure}

All 7 observed candidates displayed line emission, notably the 2.059 $\mu$m \ion{He}{i} line in each spectrum. Five of the spectra display the 2.081 $\mu$m \ion{C}{iv} line and the \ion{C}{iii} line at 2.122 $\mu$m. The individual spectra are shown in Figure~\ref{fig:specs}, and a coadded composite spectrum of all 7 objects is included in Figure~\ref{fig:coadd}. Coordinates and \textit{2MASS} magnitudes for these objects are listed in Table~\ref{tab:data}. 

\begin{table*}
\centering
\caption{New emission-line objects from this work. Object IDs are unrelated to RA and Dec.\label{tab:data}}
\begin{tabular}{lccccrrr}
\hline
\textbf{Object} & \textbf{RA}  & \textbf{Dec} & \textbf{$l$} & \textbf{$b$} & \textbf{$J$} & \textbf{$H$} & \textbf{$K$} \\
\hline
0576-1674 & 09:15:20.11 & -49:53:30.0 & 271.51 & -0.68 & 15.37 & 14.81 & 14.58 \\
0580-18E8 & 09:16:55.04 & -50:10:30.9 & 271.89 & -0.70 & 15.63 & 15.13 & 14.64 \\
0583-214C & 09:21:51.99 & -50:21:58.7 & 272.58 & -0.27 & 16.14 & 15.19 & 14.67 \\
0583-114B & 09:22:09.39 & -50:11:37.6 & 272.49 & -0.12 & 15.71 & 14.80 & 14.56 \\
0683-2EFA & 10:29:07.12 & -58:15:30.2 & 285.11 & -0.40 & 15.30 & 14.62 & 14.32 \\
0728-34FF & 11:07:02.78 & -61:09:01.6 & 290.68 & -0.79 & 14.68 & 14.45 & 14.28 \\
0914-5728 & 14:20:32.01 & -60:05:46.4 & 313.83 & 0.88 & 16.10 & 14.88 & 14.39 \\
\hline
\multicolumn{8}{c}{\textbf{Objects from \citetalias{2012AJ....143..149S} and \citetalias{2015MNRAS.452.2858K} (re-)classified as [WC]s}}\\
\hline
1023-63L & 15:52:09.48 & -54:17:14.5 & 327.39 & -0.23 & 16.13 & 15.06 & 14.37 \\
1626-4FC8 & 19:06:33.66 & +09:07:20.8 & 42.77 & 0.82 & 15.59 & 14.86 & 13.89 \\
\hline
\end{tabular}
\end{table*}

\begin{figure}
\includegraphics[width=\columnwidth]{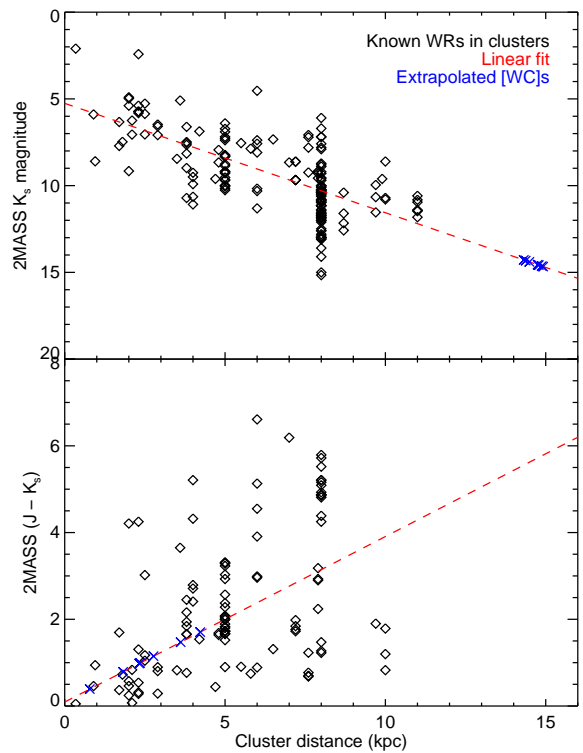}
\caption{Distance vs \textit{2MASS} $K_s$ (top) and $J-K_s$ (bottom) for classical WR stars in clusters. A linear fit is shown as the red dashed line, and the blue X's are extrapolated distances in each case for the 7 objects presented here. The severe mismatch in the two extrapolated distances demonstrates conclusively that these are not classical WR stars.}
\label{fig:clust}
\end{figure}

We have already noted above that objects which have very strong \ion{C}{iv} emission and are very blue ($J-K_s < 2$) can be classical WR stars only if $K_s < 12-13$. The seven candidate stars chosen for this study are $\sim1.5-2$ magnitudes fainter. These seven objects fall squarely in the region of the CMD occupied exclusively by PNe. Classical WR stars are far too luminous to be as faint in $K$-band as these seven candidates for the $J-K_s$ colours we observe. To demonstrate this further, in Figure~\ref{fig:clust} we show plots of distance vs $K_s$ and $J-K_s$ for classical WR stars in clusters. We then extrapolate a distance for the seven objects presented here based on a simple linear fit in each case. The two sets of distances display a severe mismatch, clear evidence that these objects cannot be classical WR stars.

\begin{figure}
\includegraphics[width=\columnwidth]{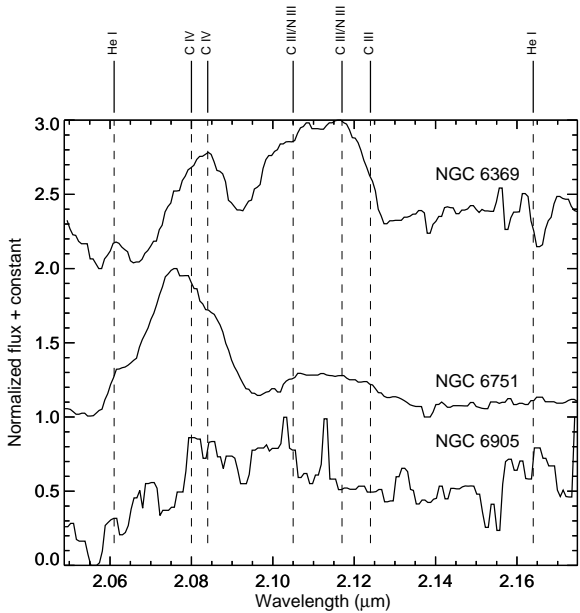}
\caption{$K$-band spectra of NGC 6369, NGC 6751, and NGC6905, the three [WC]s in Figure~7 of \citet{2012ApJS..200....3B}, from observations with NIRI on Gemini. These spectra are newly reduced for this paper, using the \textsc{gemini.niri} package in \textsc{iraf}. The [WC] spectra display very little Br-$\gamma$ emission at 2.16 $\mu$m.}
\label{fig:gemini}
\end{figure}

One might expect a CSPNe spectrum to show a strong nebular 2.162~$\mu$m Br-$\gamma$ line, but in [WC] stars this line is weak or nonexistent. Figure~\ref{fig:gemini} shows $K$-band spectra of three [WC]s from \citet[fig.~7]{2012ApJS..200....3B}. Only one of the [WC]s contains a significant Br-$\gamma$ lines, and it is weak compared to the strong WR lines. Could these 7 objects instead be very hot O stars? The answer is no, as is made abundantly clear by the $K$-band spectra of hot O stars in \citet{2005ApJS..161..154H}. For those stars, in every case the 2.112~$\mu$m \ion{He}{i}/\ion{N}{iii} doublet is stronger and/or broader than the 2.081 \ion{C}{iv} line, either in emission or absorption, while in the 7 objects presented here, this 2.112~$\mu$m line is either extremely weak or entirely absent.

\begin{figure}
\includegraphics[width=\columnwidth]{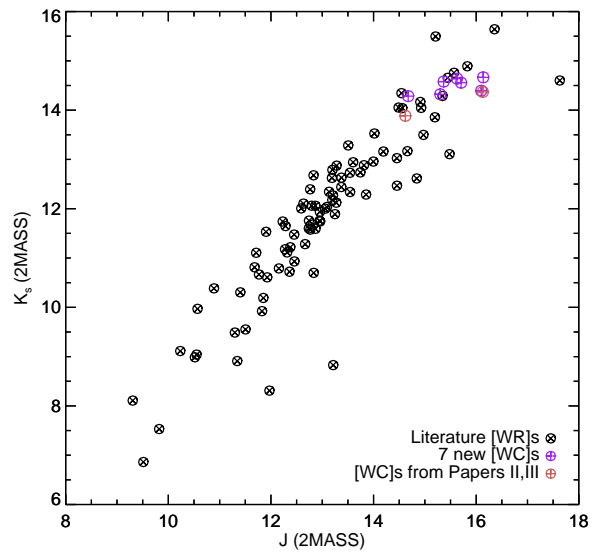}
\caption{$J$ vs $K_s$ magnitude for all [WR]s in the literature ([WC], [WN], and indeterminate [WR]), the two [WC]s from Papers I \& II, and the 7 new [WC]s described in this paper. These 7 objects are among the faintest [WR] stars ever identified.}
\label{fig:jk}
\end{figure}

\begin{figure}
\includegraphics[width=\columnwidth]{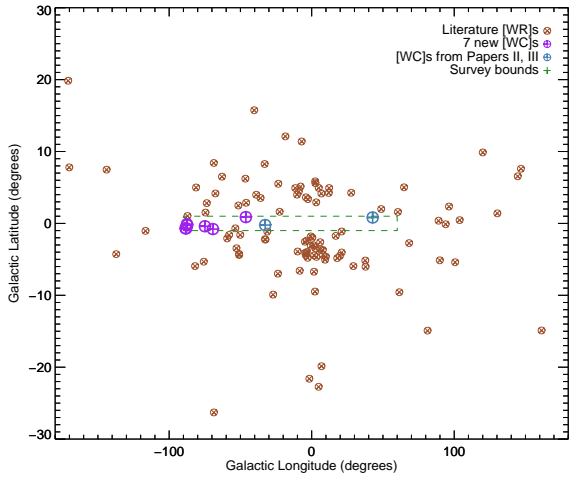}
\caption{A projection on the sky of [WR]s of all types from the literature, the new [WC]s described in this paper, and [WC]s from other papers associated with this survey (see section~\ref{sec:oth}). The survey bounds are shown with a dashed line; only one of the [WR]s from the literature lies within the survey, and only 4 lie within $b=\pm1^{\circ}$.}
\label{fig:project}
\end{figure}

Figure~\ref{fig:jk} shows that these 7 [WC]s are among the faintest [WR] stars ever identified. They also are drawn from a survey of the Galactic plane, a region heavily undersampled by the existing CSPNe literature, as shown in Figure~\ref{fig:project}. These observations confirm the hypothesis that the given selection criteria can reliably select [WC] stars, even in crowded fields in the Galactic plane.

The ionization potentials of \ion{C}{iv}, \ion{He}{i}, and \ion{He}{ii} are $\approx 64.5$, 24.5, and 54.4 eV respectively, corresponding to minimum source temperatures of 64, 25, and 54 kK respectively \citep{NIST_ASD}. The presence of \ion{C}{iv} lines in the 7 spectra shown here implies a stellar temperature of $\ga65$ kK, but not much greater due to the lack of strong \ion{He}{ii} \citep[see also][figs.~5-8]{2011MNRAS.418..705K}.

\subsection{Other [WC]s from this survey}\label{sec:oth}
\begin{figure}
\includegraphics[width=\columnwidth]{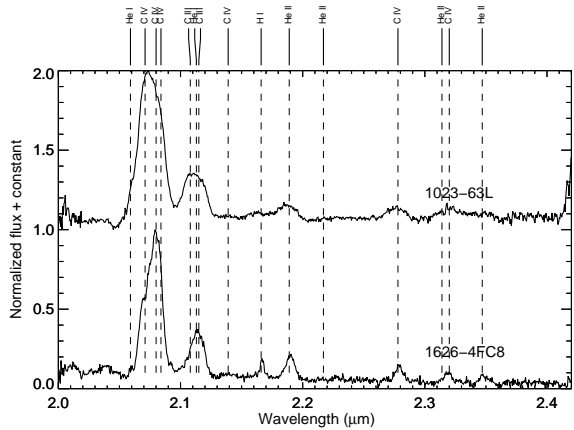}
\caption{Spectra for two previously-identified [WC]s, 1023-63L from \citetalias{2012AJ....143..149S} and 1626-4FC8 from \citetalias{2015MNRAS.452.2858K}. 1023-63L was originally classified as a WC7: in \citetalias{2012AJ....143..149S}. Only the $K_s$-band spectra for these objects are included, for comparison with the rest of these observations.}
\label{fig:old}
\end{figure}

One object originally identified as a [WC] \citepalias[1626-4FC8 in][]{2015MNRAS.452.2858K} lies in the faint, blue region of Figure~\ref{fig:kjk}, as does object 1023-63L from \citetalias{2012AJ....143..149S} which was originally classified as a WC7:. In light of the 7 new objects we describe in this paper, we maintain that 1023-63L should be re-classified as a [WC7:]. These two [WC]s are included in Table~\ref{tab:data}, and their spectra in Figure~\ref{fig:old}.

\section{Massive CSPNe}\label{sec:mass}
\begin{figure}
\includegraphics[width=\columnwidth]{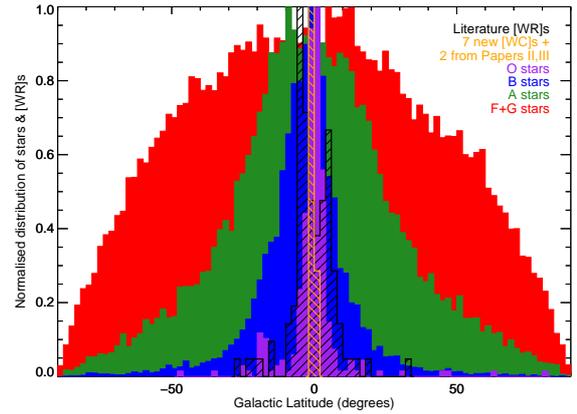}
\caption{The normalised distribution of stars and [WR]s in Galactic latitude. The progression from F and G stars (red) to A stars (green), B stars (blue), and then to O stars (purple), shows increased clustering toward the Galactic plane as a function of stellar mass. The [WR] star distribution is also clustered tightly near the Galactic plane, supporting the hypothesis of \citet{2011A&A...526A...6W} that [WR]s are likely to arise from more-massive progenitors than H-rich CSPNe. However, the literature shows extreme undersampling within $1^{\circ}$ of the Galactic plane, where the most massive CSPNe are likely to reside. The 7 new [WC]s presented in this paper (as well as two other from the same survey of the Galactic plane) begin to fill the gap. B, A, and G stars taken from Hipparcos \citep{1997A&A...323L..49P}, Gliese \citep{1995yCat.5070....0G}, and Yale BSC \citep{1995yCat.5050....0H}.}
\label{fig:galdist}
\end{figure}

Analyses in \citet[see Figure~1]{2011A&A...526A...6W} show that H-poor CSPNe (including [WR] stars) cluster more strongly toward the Galactic plane than H-rich CSPNe. Figure~\ref{fig:galdist} shows the distribution in Galactic latitude of [WR] stars, overlaid on a progression of stars from \textit{Hipparcos} \citep{1997A&A...323L..49P}, \textit{Gliese} \citep{1995yCat.5070....0G}, and \textit{Yale BSC} \citep{1995yCat.5050....0H}, from O ($\ge15.5$ M\sun, from \citealt{2005A&A...436.1049M}) $\rightarrow$ B ($2-18$ M\sun, from \citealt{1981A&AS...46..193H,2014A&A...566A...7N}) $\rightarrow$ A ($1.5-2.5$ M\sun, from \citealt{1981A&AS...46..193H,2013ApJ...771...40B}) $\rightarrow$ F \& G ($0.2-2$ M\sun, from \citealt{1981A&AS...46..193H,2013ApJ...771...40B}). It is clear that the more massive stars cluster more strongly toward the Galactic plane, which suggests that H-poor CSPNe evolve from more massive progenitor stars than H-rich PNe \citep{2011A&A...526A...6W}. 

However, Figure~\ref{fig:galdist} also demonstrates the observational gap in the literature for [WR] stars within $\pm1^{\circ}$ of the Galactic plane, where the proportion of [WR] stars should be the highest. These 7 new [WC]s lie directly in this undersampled region, as evidence that this region contains large numbers of [WR] stars yet unidentified. A future survey for Galactic plane [WR] stars must be carried out, to constrain the properties of this evolutionary stage between the post-AGB phase and the white dwarf (WD) phase. The presence (or absence) and relative strengths of \ion{He}{i}, \ion{He}{ii}, \ion{C}{iii}, and \ion{C}{iv} emission lines in such spectra can be used to gauge the temperature of this pre-WD population, and provide insight into the high-mass end of the Initial-Final Mass Relation \citep{2008ApJ...676..594K}.

\section{Conclusions}\label{sec:awk}

The 7 emission-line objects presented here were selected to be nearby, faint [WC]s. The 100 per cent success rate of our mini-survey for these objects shows that the narrowband color criteria of this Galactic Plane survey are powerful tools for identifying [WC] CSPNe, in addition to the already-proven success for finding classical WR stars and less-exotic PNe. There are thousands more candidates, similar to the seven we have described here, in our survey catalogs, awaiting spectrographic follow-up. Spectrographic follow-up of these candidates will result in hundreds of confirmed [WC] stars in the Galactic plane, from which rigorous statistical analysis of these very interesting objects will be made possible.

\section*{Acknowledgements}
This work is based on observations made with ESO Telescopes at the La Silla Paranal Observatory under programme ID 290.D-5121(A). This research makes use of data products from the Two-Micron All-Sky Survey and the NASA/IPAC Infrared Science Archive. GK would like to acknowledge extensive hands-on assistance from Kathleen Labrie in reducing the Gemini spectra. MS and GK gratefully acknowledge support from Hilary and Ethel Lipsitz, longtime friends of the Astrophysics department at the American Museum of Natural History.

%%%%%%%%%%%%%%%%%%%%%%%%%%%%%%%%%%%%%%%%%%%%%%%%%%

%%%%%%%%%%%%%%%%%%%% REFERENCES %%%%%%%%%%%%%%%%%%

% The best way to enter references is to use BibTeX:

\bibliographystyle{mnras}
%\bibliography{vlt} % if your bibtex file is called example.bib

%%%%%%%%%%%%%%%%%%%%%%%%%%%%%%%%%%%%%%%%%%%%%%%%%%

%%%%%%%%%%%%%%%%% APPENDICES %%%%%%%%%%%%%%%%%%%%%

\appendix

\section{Finder Charts}
Presented here are finder charts for the 7 new objects described in this paper.

\begin{figure*}
\includegraphics[width=0.75\linewidth]{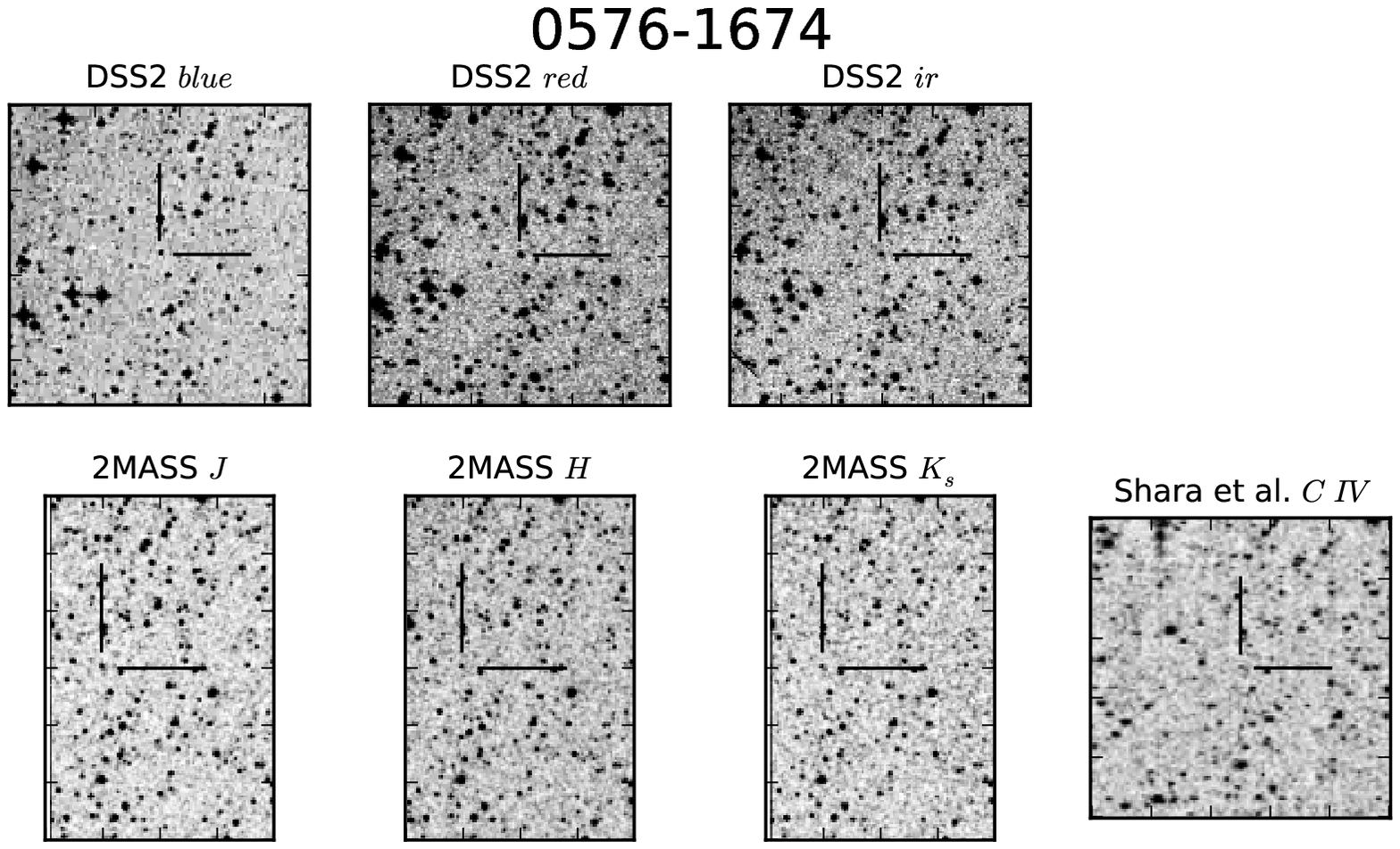}
\includegraphics[width=0.75\linewidth]{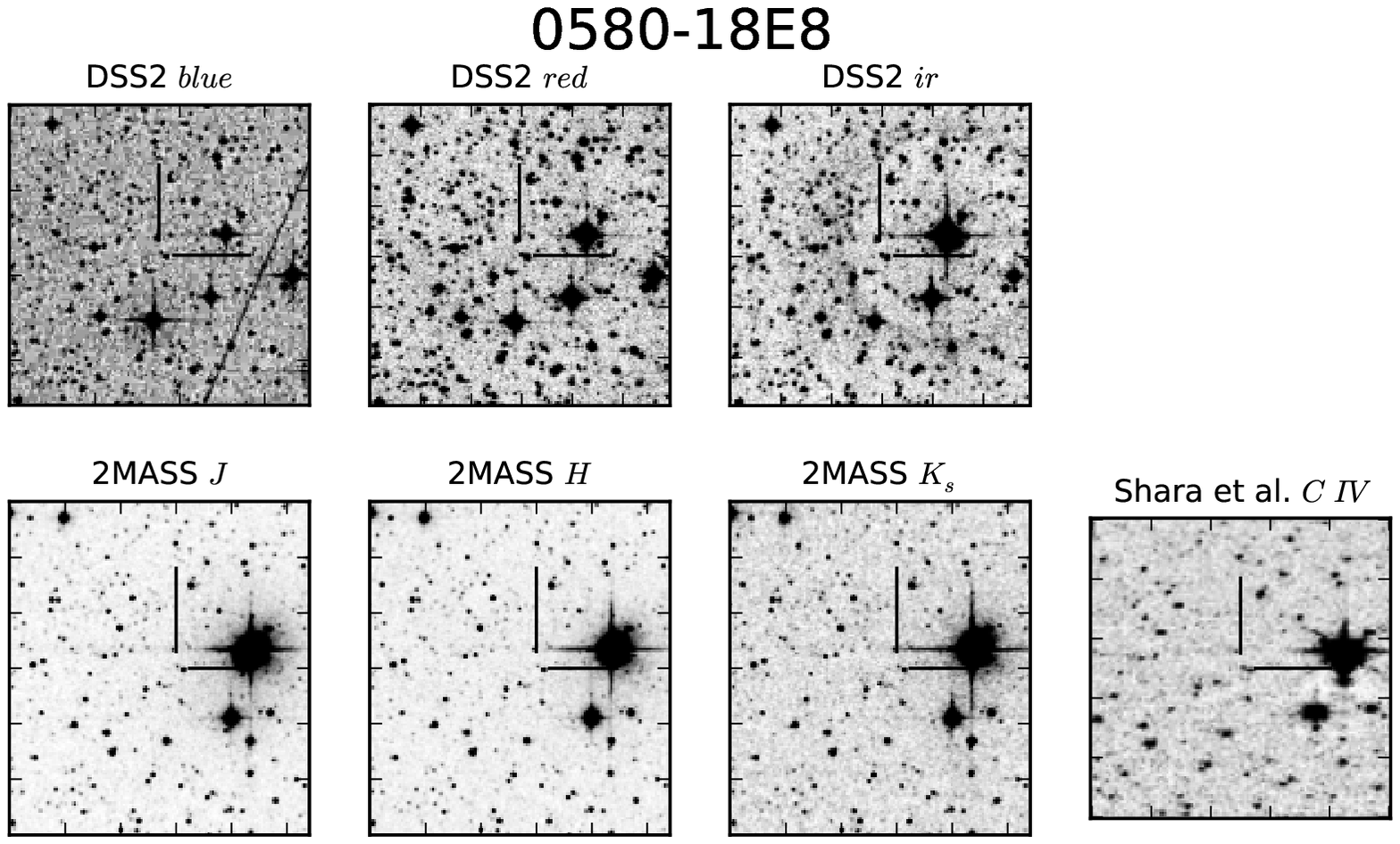}
\includegraphics[width=0.75\linewidth]{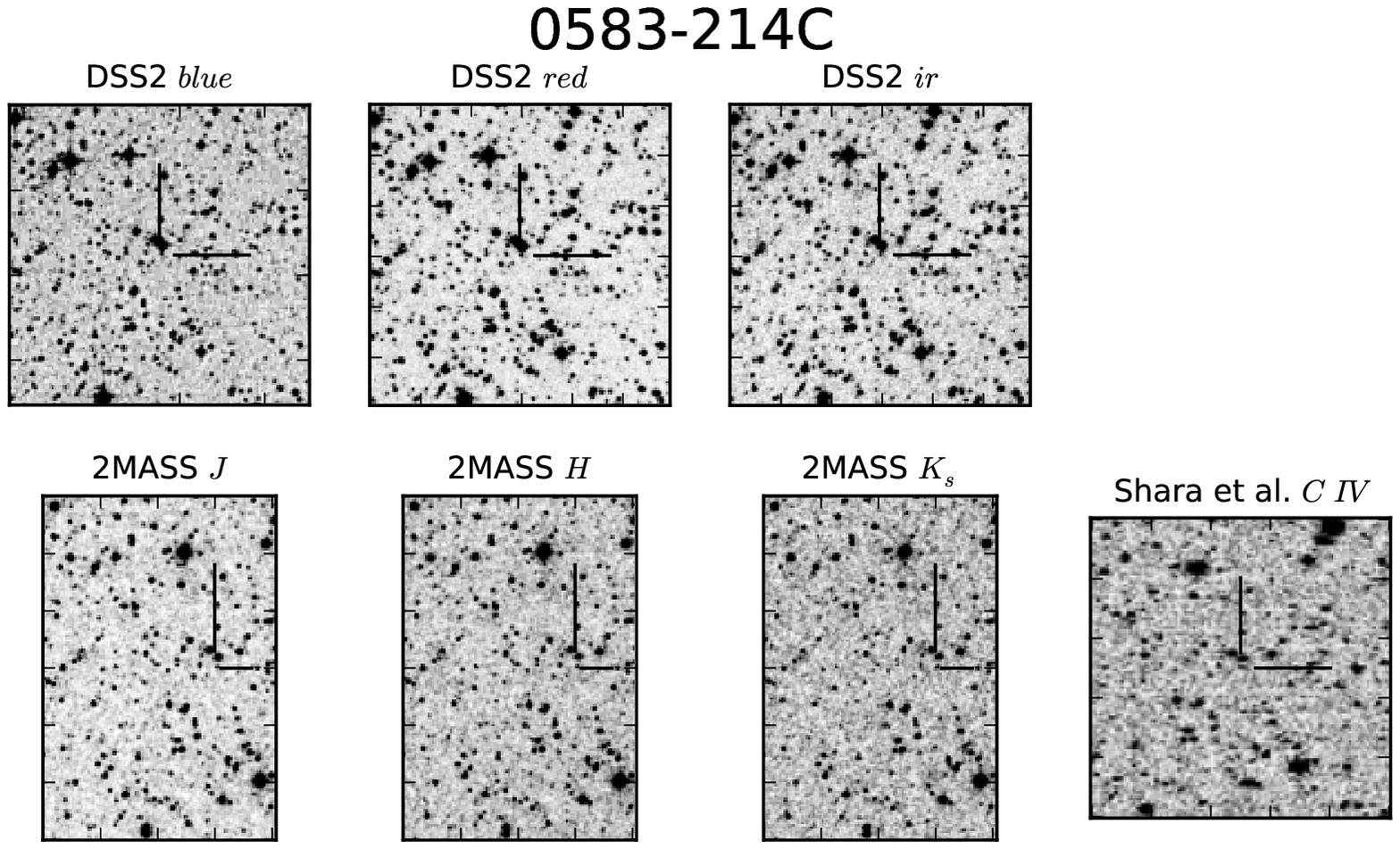}
\caption{Finder charts for the new objects presented in this paper. Each image is 5 arcminutes across, with North up and East to the left.}
\label{fig:fc}
\end{figure*}
\begin{figure*}
\ContinuedFloat
\includegraphics[width=0.75\linewidth]{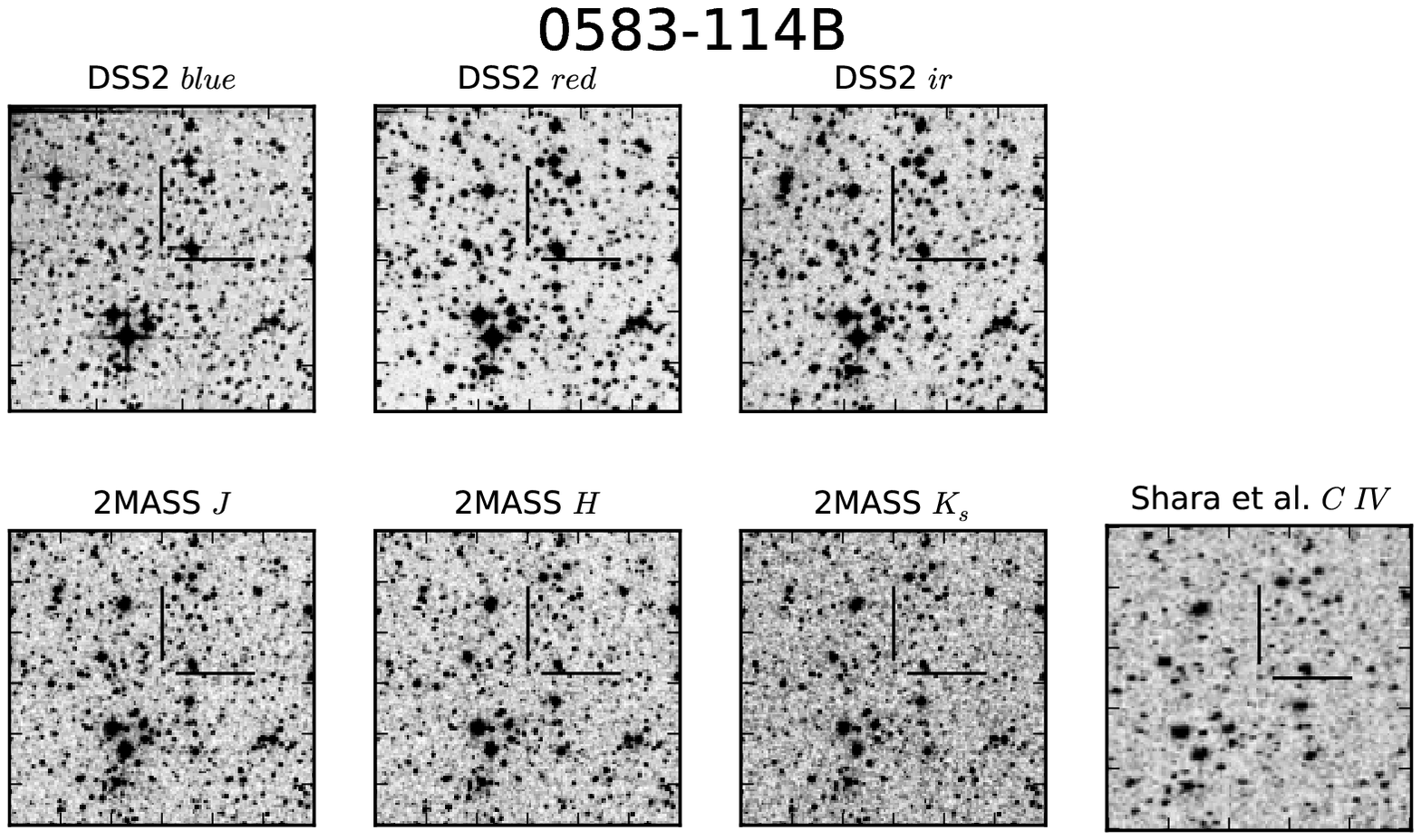}
\includegraphics[width=0.75\linewidth]{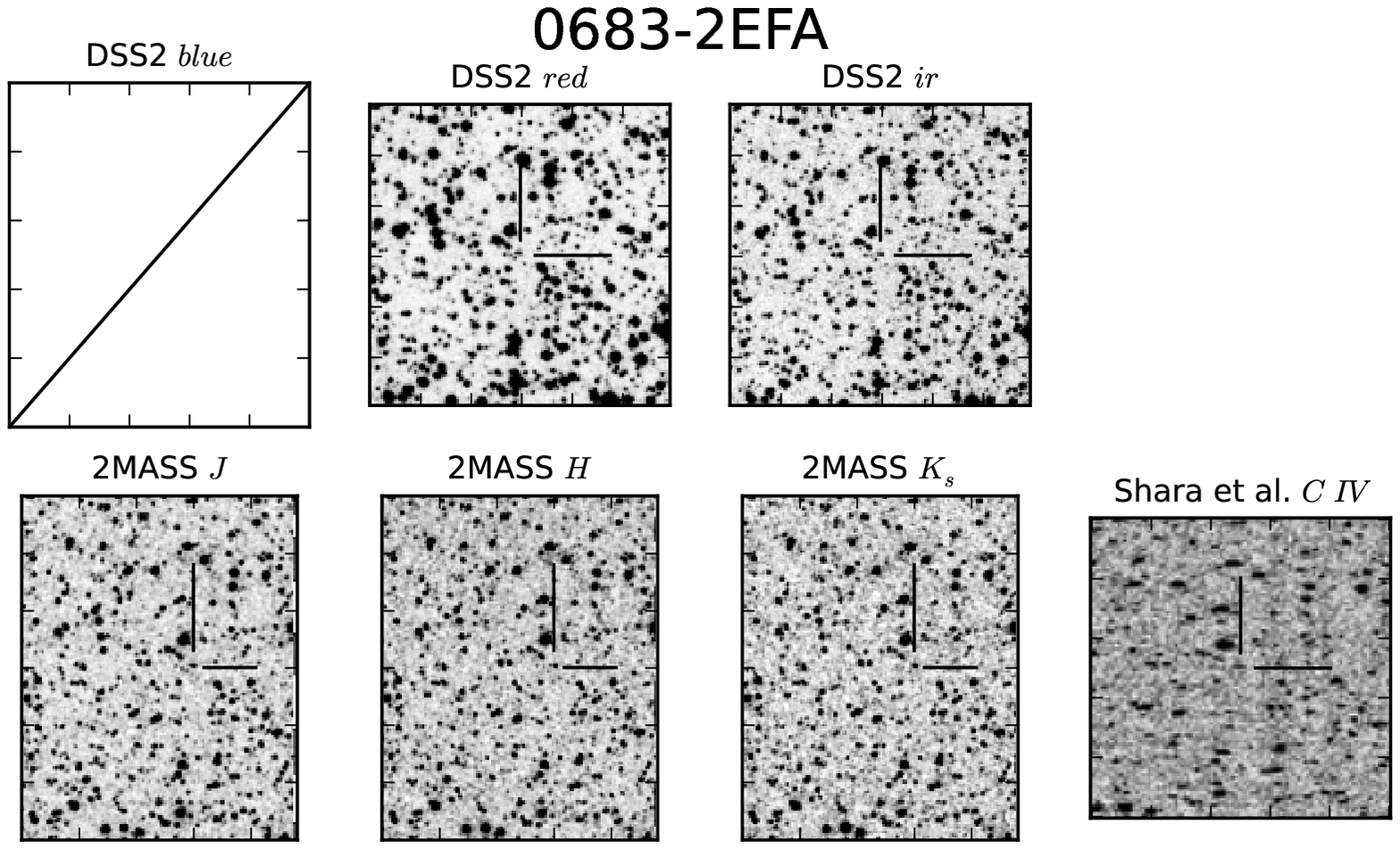}
\includegraphics[width=0.75\linewidth]{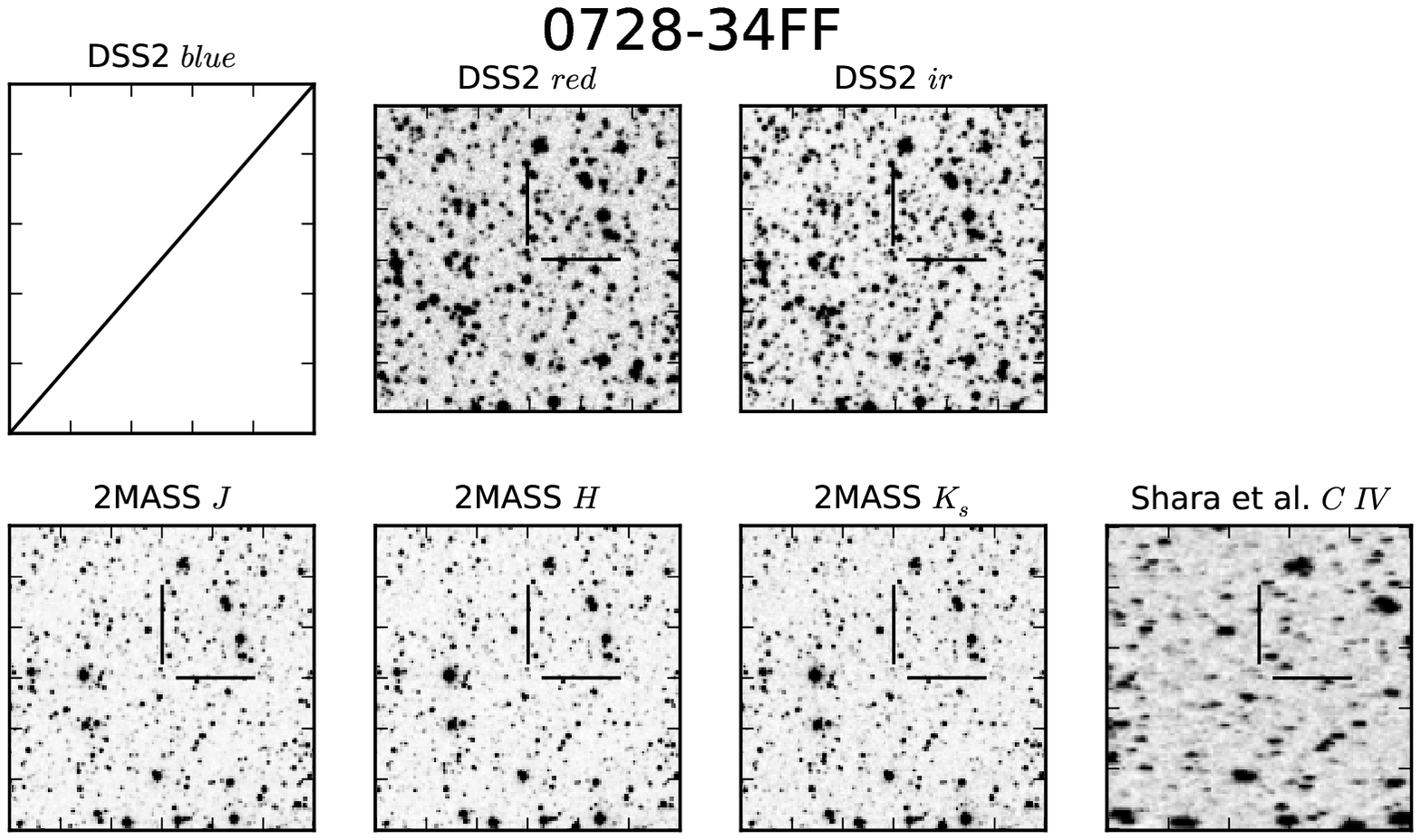}
\caption{\textit{(cont'd)} Finder charts for the new objects presented in this paper. Each image is 5 arcminutes across, with North up and East to the left.}
\label{fig:fc_1}
\end{figure*}
\begin{figure*}
\ContinuedFloat
\includegraphics[width=0.75\linewidth]{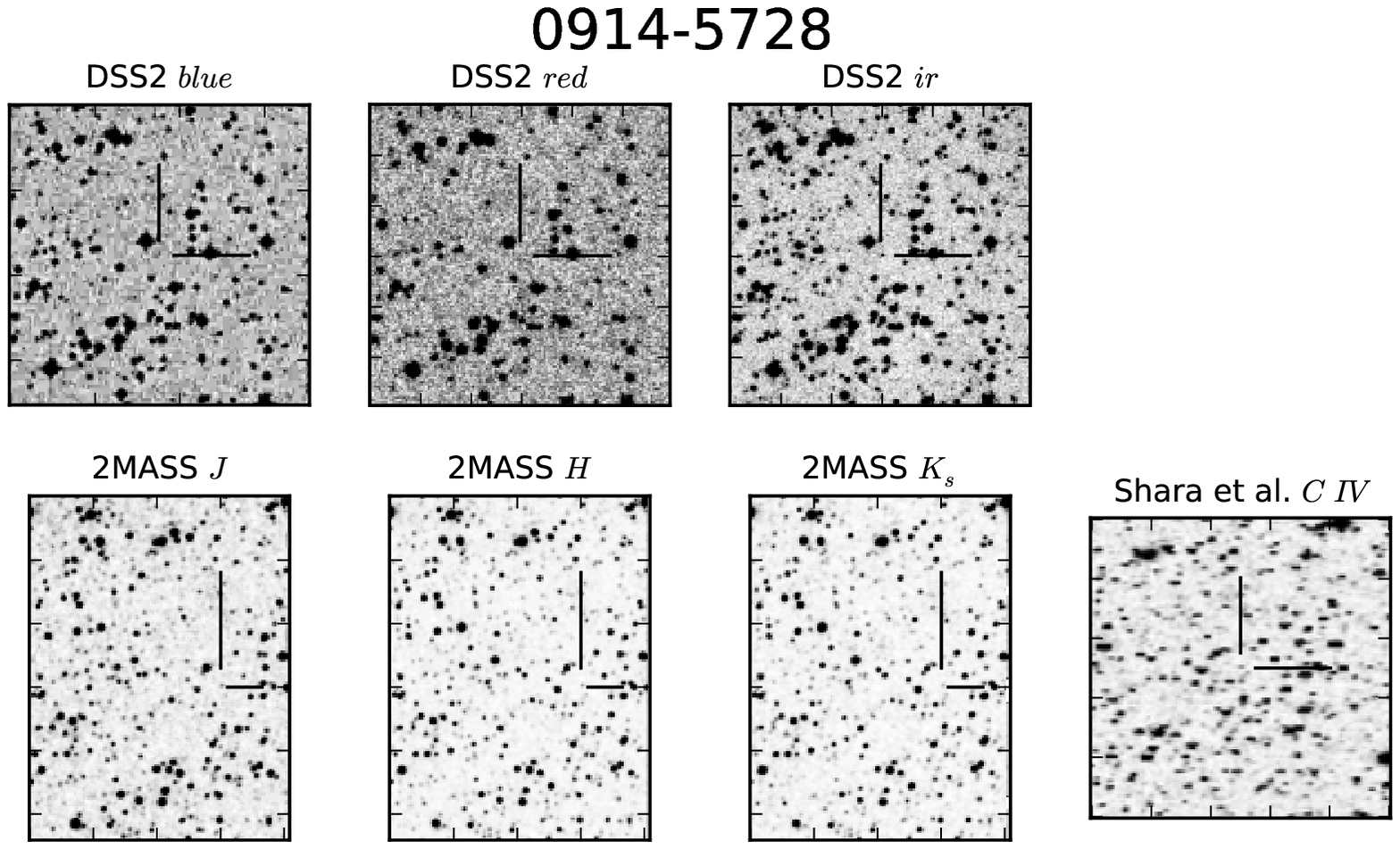}
\caption{\textit{(cont'd)} Finder charts for the new objects presented in this paper. Each image is 5 arcminutes across, with North up and East to the left.}
\label{fig:fc_2}
\end{figure*}

%%%%%%%%%%%%%%%%%%%%%%%%%%%%%%%%%%%%%%%%%%%%%%%%%%

% Don't change these lines
\bsp	% typesetting comment
\label{lastpage}
\end{document}